\documentclass{article} 
\usepackage{nips13submit_e,times}
\usepackage{hyperref}
\usepackage{url}

\usepackage{alltt}
\usepackage[latin1]{inputenc}
\usepackage{mathtools}
\usepackage{graphicx}
\usepackage{epstopdf}
\usepackage{multirow}
\usepackage[bottom]{footmisc}

\usepackage{algorithm}
\usepackage[noend]{algorithmic}

\usepackage{amsmath}
\usepackage{amsfonts}
\usepackage{amssymb}
\usepackage{dsfont}
\usepackage{float}
\usepackage{url}
\usepackage{textcomp}
\usepackage{cite}
\usepackage{pdfpages}

\newcommand{\bs}[1]{\ensuremath{\boldsymbol{#1}}}
\newcommand{\mc}[1]{\ensuremath{\mathcal{#1}}}

\newcommand{\proj}[1]{P^{\left(\mc{#1}\right)}}
\newcommand{\pth}[1]{\left(#1\right)}

\newcommand{\bigOO}{O}
\newcommand{\inv}[1]{\left( #1 \right)^{-1}}

\newcommand{\BlackBox}{\rule{1.5ex}{1.5ex}}
\newenvironment{proof}{\par\noindent{\bfseries\upshape
  Proof\ }}{\hfill\BlackBox\\[2mm]}

\newtheorem{theorem}{Theorem}

\newtheorem{lemma}[theorem]{Lemma}

\newtheorem{problem}{Problem}

\title{\Large A Fast Greedy Algorithm for \\ Generalized Column Subset Selection}

\author{
Ahmed K. Farahat, Ali Ghodsi, and Mohamed S. Kamel\\
University of Waterloo, Waterloo, Ontario, N2L 3G1, Canada\\
\texttt{\{afarahat,aghodsib,mkamel\}@uwaterloo.ca}
}
\nipsfinalcopy 

\begin{document}

\maketitle

\begin{abstract}
This paper defines a generalized column subset selection problem which is concerned with the selection of a few columns from a source matrix $A$ that best approximate the span of a target matrix $B$. The paper then proposes a fast greedy algorithm for solving this problem and draws connections to different problems that can be efficiently solved using the proposed algorithm.
\end{abstract}

\section{Generalized Column Subset Selection}\label{Sec:CSS}
The Column Subset Selection (CSS) problem can be generally defined as the selection of a few columns from a data matrix that best approximate its span \cite{Frieze98-Rnd, Drineas06-Cols, Boutsidis08-Clust, Boutsidis09a-CSS, Boutsidis11a-NearOpt,  Civril12-CSS-Sparse}. 
We extend this definition to the generalized problem of selecting a few columns from a source matrix to approximate the span of a target matrix.
The generalized CSS problem can be formally defined as follows:
\begin{problem} {\bf (Generalized Column Subset Selection)} \label{Pr:GenCSSNew} \label{Pr:FS}
    Given a source matrix $A\in\mathbb{R}^{m\times n}$, a target matrix $B\in\mathbb{R}^{m\times r}$ and an integer $l$, find a subset of columns $\mc{L}$ from $A$ such that $|\mc{L}| =l$ and
    \begin{displaymath}
        \mc{L}={\arg\min}_{\mc{S}}\:\|B-\proj{S}B\|_{F}^{2},
    \end{displaymath} where $\mc{S}$ is the set of the indices of the candidate columns from $A$, $\proj{S}\in\mathbb{R}^{m\times m}$ is a projection matrix which projects the columns of $B$ onto the span of the set $\mc{S}$ of columns, and $\mc{L}$ is the set of the indices of the selected columns from $A$.
\end{problem}

The CSS criterion $\mathbf{F}\left(\mc{S}\right)=\|B-\proj{S}B\|_{F}^{2}$ represents the sum of squared errors between the target matrix $B$ and its rank-$l$ approximation $\proj{S}B$ . In other words, it calculates the Frobenius norm of the residual matrix $F=B-\proj{S}B$. Other types of matrix norms can also be used to quantify the reconstruction error \cite{Boutsidis09a-CSS, Boutsidis11a-NearOpt}. The present work, however, focuses on developing algorithms that minimize the Frobenius norm of the residual matrix. The projection matrix $\proj{S}$ can be calculated as
$\proj{S}=A_{:\mc{S}} \inv{A_{:\mc{S}}^{T}A_{:\mc{S}}} A_{:\mc{S}}^{T} \:,$ where $A_{:\mc{S}}$ is the sub-matrix of $A$ which consists of the columns corresponding to $\mc{S}$. It should be noted that if $\mc{S}$ is known, the term $\inv{A_{:\mc{S}}^{T}A_{:\mc{S}}} A_{:\mc{S}}^{T}B$ is the closed-form solution of least-squares problem  $T^{*}={\arg\min}_T\left\Vert B-A_{:\mc S}T\right\Vert _{F}^{2}$.

\section{A Fast Greedy Algorithm for Generalized CSS}
Problem \ref{Pr:GenCSSNew} is a combinatorial optimization problem whose optimal solution can be obtained in $O\pth{\max\pth{n^lmrl, n^lml^2}}$. In order to approximate this optimal solution, we propose a fast greedy algorithm that selects one column from $A$ at a time. The greedy algorithm is based on a recursive formula for the projection matrix $P^{(\mc{S})}$ which can be derived as follows.


\begin{lemma}  \label{Lm:Proj}
    Given a set of columns $\mc{S}$. For any $\mc{P} \subset \mc{S}$,
        $\proj{S}=P^{\left(\mc{P}\right)}+R^{\left(\mc{R}\right)}\:,     $
    where $R^{\left(\mc{R}\right)} = E_{:\mc{R}} \inv{E_{:\mc{R}}^{T}E_{:\mc{R}}} E_{:\mc{R}}^{T}$ is a projection matrix which
    projects the columns of $E = A - P^{\left(\mc{P}\right)} A$
    onto the span of the subset $\mc{R} = \mc{S}\setminus \mc{P}$ of
    columns.
\end{lemma}

\proof Define $D=A_{:\mc{S}}^{T}A_{:\mc{S}}$. 
The projection matrix $\proj{S}$ can
be written as $\proj{S}=A_{:\mc{S}}D^{-1}A_{:\mc{S}}^{T}$. Without loss of generality, the columns and rows of $A_{:\mc{S}}$
and $D$ can be rearranged such that
the first sets of rows and columns correspond to $\mc{P}$. 
Let $S=D_{\mc{R}\mc{R}}-D_{\mc{P}\mc{R}}^{T}D_{\mc{P}\mc{P}}^{-1}D_{\mc{P}\mc{R}}$
be the Schur complement \cite{lutkepohl1996handbook} of
$D_{\mc{P}\mc{P}}$ in $D$, where $D_{\mc{P}\mc{P}}=A_{:\mc{P}}^{T}A_{:\mc{P}}$,
$D_{\mc{P}\mc{R}}=A_{:\mc{P}}^{T}A_{:\mc{R}}$ and $D_{\mc{R}\mc{R}}=A_{:\mc{R}}^{T}A_{:\mc{R}}$. Using the block-wise inversion formula
\cite{lutkepohl1996handbook}, $D^{-1}$ can be calculated as
\begin{displaymath}
D^{-1}= \left[\begin{array}{cc}
        D_{\mc{P}\mc{P}}^{-1}+D_{\mc{P}\mc{P}}^{-1}D_{\mc{P}\mc{R}}S^{-1}D_{\mc{P}\mc{R}}^{T}D_{\mc{P}\mc{P}}^{-1} & -D_{\mc{P}\mc{P}}^{-1}D_{\mc{P}\mc{R}}S^{-1}\\
        -S^{-1}D_{\mc{P}\mc{R}}^{T}D_{\mc{P}\mc{P}}^{-1}
        & S^{-1}\end{array}\right]
\end{displaymath}
Substituting with $A_{:\mc{S}}$ and $D^{-1}$ in $\proj{S}=A_{:\mc{S}}D^{-1}A_{:\mc{S}}^{T}$, the projection matrix can be simplified to
    \begin{equation} \label{Eq:ProjP2}
        \begin{split}
        \proj{S}=A_{:\mc{P}}D_{\mc{P}\mc{P}}^{-1}A_{:\mc{P}}^{T}
        +\left(A_{:\mc{R}}-A_{:\mc{P}}D_{\mc{P}\mc{P}}^{-1}D_{\mc{P}\mc{R}}\right)S^{-1}\left(A_{:\mc{R}}^{T}-D_{\mc{P}\mc{R}}^{T}D_{\mc{P}\mc{P}}^{-1}A_{:\mc{P}}^{T}\right) \:.
        \end{split}
    \end{equation}
The first term of the right-hand side is the projection matrix $P^{\left(\mc{P}\right)}$ which projects vectors onto the span of the subset $\mc{P}$ of columns.
The second term can be simplified as follows. Let $E$ be an $m
\times n$ residual matrix which is calculated as:
$E=A-P^{\left(\mc{P}\right)}A$. The sub-matrix $E_{:\mc{R}}$ can be expressed as
\begin{displaymath}
   E_{:\mc{R}}=A_{:\mc{R}}-P^{\left(\mc{P}\right)}A_{:\mc{R}} = A_{:\mc{R}}-A_{:\mc{P}}\pth{A_{:\mc{P}}^{T}A_{:\mc{P}}}^{-1}A_{:\mc{P}}^{T}A_{:\mc{R}}=A_{:\mc{R}}-A_{:\mc{P}}D_{\mc{P}\mc{P}}^{-1}D_{\mc{P}\mc{R}} \:.
\end{displaymath}
Since projection matrices are idempotent, then $P^{\left(\mc{P}\right)}P^{\left(\mc{P}\right)}=P^{\left(\mc{P}\right)}$ and 
\begin{displaymath}
       E_{:\mc{R}}^{T}E_{:\mc{R}} = \pth{A_{:\mc{R}}-P^{\left(\mc{P}\right)}A_{:\mc{R}}}^T \pth{A_{:\mc{R}}-P^{\left(\mc{P}\right)}A_{:\mc{R}}}
       =A_{:\mc{R}}^TA_{:\mc{R}}-A_{:\mc{R}}^TP^{\left(\mc{P}\right)}A_{:\mc{R}}\:.
\end{displaymath}
Substituting with $P^{\left(\mc{P}\right)}=A_{:\mc{P}}\pth{A_{:\mc{P}}^{T}A_{:\mc{P}}}^{-1}A_{:\mc{P}}^T$ gives
\begin{displaymath}
    \begin{split}
       E_{:\mc{R}}^{T}E_{:\mc{R}}    =A_{:\mc{R}}^TA_{:\mc{R}}-A_{:\mc{R}}^TA_{:\mc{P}}\pth{A_{:\mc{P}}^{T}A_{:\mc{P}}}^{-1}A_{:\mc{P}}^TA_{:\mc{R}}= D_{\mc{R}\mc{R}}-D_{\mc{P}\mc{R}}^{T}D_{\mc{P}\mc{P}}^{-1}D_{\mc{P}\mc{R}} = S \:.
    \end{split}
\end{displaymath}
Substituting $\pth{A_{:\mc{P}}D_{\mc{P}\mc{P}}^{-1}A_{:\mc{P}}^{T}}$, $\pth{A_{:\mc{R}}-A_{:\mc{P}}D_{\mc{P}\mc{P}}^{-1}D_{\mc{P}\mc{R}}}$ and $S$ with $P^{\left(\mc{P}\right)}$, $E_{:\mc{R}}$ and $E_{:\mc{R}}^{T}E_{:\mc{R}}$ respectively, Equation (\ref{Eq:ProjP2}) can be expressed as
\begin{displaymath}
        \begin{split}
        \proj{S}=\proj{P} + E_{:\mc{R}}\pth{E_{:\mc{R}}^{T}E_{:\mc{R}}}^{-1}E_{:\mc{R}}^T\:.
        \end{split}
\end{displaymath}
The second term is the projection matrix $R^{\left(\mc{R}\right)}$ which projects vectors onto the span of $E_{:\mc{R}}$. This proves that $\proj{S}$ can be written in terms of $P^{\left(\mc{P}\right)}$ and $R$ as $\proj{S}=P^{\left(\mc{P}\right)}+R^{\left(\mc{R}\right)}$
\hfill\BlackBox

Given the recursive formula for $\proj{S}$, the following theorem derives a recursive formula for $\mathbf{F}\left(\mathcal{S}\right)$.
\begin{theorem}\label{Th:RecF}
    Given a set of columns $\mc{S}$. For any $\mc{P} \subset \mc{S}$,
$      \mathbf{F}\left(\mathcal{S}\right)=\mathbf{F}\left(\mathcal{P}\right)-\left\Vert R^{\left(\mathcal{R}\right)}F\right\Vert _{F}^{2}
 \:,$
    where $F = B - P^{\left(\mc{P}\right)}B$ and
    $R^{\left(\mc{R}\right)}$ is a projection matrix which
    projects the columns of $F$
    onto the span of the subset $\mc{R} = \mc{S}\setminus \mc{P}$ of columns of $E=A - P^{\left(\mc{P}\right)}A$
\end{theorem}
\proof By definition, $\mathbf{F}\left(\mathcal{S}\right)=\left\Vert B-P^{\left(\mathcal{S}\right)}B\right\Vert _{F}^{2}$. Using Lemma \ref{Lm:Proj}, $P^{\left(\mathcal{S}\right)}B=P^{\left(\mathcal{P}\right)}B+R^{\left(\mathcal{R}\right)}B$. The term $R^{\left(\mathcal{R}\right)}B$ is equal to $R^{\left(\mathcal{R}\right)}F$ as
$E_{:\mc{R}}^{T}B = E_{:\mc{R}}^{T}F$. To prove that,
multiplying $E_{:\mc{R}}^{T}$ by $F = B -
P^{\left(\mc{P}\right)}B$ gives $   E_{:\mc{R}}^{T}F=E_{:\mc{R}}^{T}B-E_{:\mc{R}}^{T}P^{\left(\mc{P}\right)}B$. Using $E_{:\mc{R}}=A_{:\mc{R}}-P^{\left(\mc{P}\right)}A_{:\mc{R}}$,
the expression $E_{:\mc{R}}^{T}P^{\left(\mc{P}\right)}$
can be written as $        E_{:\mc{R}}^{T}P^{\left(\mc{P}\right)}=A_{:\mc{R}}^{T}P^{\left(\mc{P}\right)}-A_{:\mc{R}}^{T}P^{\left(\mc{P}\right)}P^{\left(\mc{P}\right)}$. This is equal to $0$ as $P^{\left(\mc{P}\right)}P^{\left(\mc{P}\right)}=P^{\left(\mc{P}\right)}$
(an idempotent matrix).
Substituting in $\mathbf{F}\left(\mathcal{S}\right)$ and using $F=B-P^{\left(\mathcal{P}\right)}B$ gives
\begin{equation*}
  \mathbf{F}\left(\mathcal{S}\right)=\left\Vert B-P^{\left(\mathcal{P}\right)}B-R^{\left(\mathcal{R}\right)}F\right\Vert _{F}^{2} = \left\Vert F-R^{\left(\mathcal{R}\right)}F\right\Vert _{F}^{2}
\end{equation*}
Using the relation between Frobenius norm and trace, $\mathbf{F}\left(\mathcal{S}\right)$ can be simplified to
\begin{displaymath}
  \mathbf{F}\left(\mathcal{S}\right)=\text{tr}\left(\left(F-R^{\left(\mathcal{R}\right)}F\right)^{T}\left(F-R^{\left(\mathcal{R}\right)}F\right)\right)
  =\text{tr}\left(F^{T}F-F^{T}R^{\left(\mathcal{R}\right)}F\right)=\left\Vert F\right\Vert _{F}^{2}-\left\Vert R^{\left(\mathcal{R}\right)}F\right\Vert _{F}^{2}
\end{displaymath}
Using $\mathbf{F}\left(\mathcal{P}\right)=\left\Vert F\right\Vert _{F}^{2}$ proves the theorem. \hfill\BlackBox

Using the recursive formula for $\mathbf{F}\left(\mathcal{S}\cup\{i\}\right)$ allows the development of a greedy algorithm which at iteration $t$ selects column $p$ such that
\begin{equation*}
  p={\arg\min}_i\:\mathbf{F}\left(\mathcal{S}\cup\{i\}\right)={\arg\max}_i\left\Vert P^{\pth{\left\{ i\right\}}}F\right\Vert _{F}^{2}\:.
\end{equation*}

Let $G=E^TE$ and $H=F^TE$, the objective function $\left\Vert P^{\pth{\left\{ i\right\}}}F\right\Vert _{F}^{2}$ can be simplified to
\begin{equation*}
\left\Vert E_{:i}\left(E_{:i}^{T}E_{:i}\right)^{-1}E_{:i}^{T}F\right\Vert _{F}^{2}=\text{tr}\left(F^TE_{:i}\left(E_{:i}^{T}E_{:i}\right)^{-1}E_{:i}^{T}F\right)=\frac{\left\Vert F^TE_{:i}\right\Vert ^{2}}{E_{:i}^{T}E_{:i}}=\frac{\left\Vert H_{:i}\right\Vert ^{2}}{G_{ii}}\:.
\end{equation*}
This allows the definition of the following greedy generalized CSS problem.
\begin{problem} {\bf (Greedy Generalized CSS)} \label{Pr:GreedyCSS}
    At iteration $t$, find column $p$ such that
    \begin{equation*}
        p={\arg\max}_i\hspace{1em}\frac{\left\Vert H_{:i}\right\Vert ^{2}}{G_{ii}}
    \end{equation*}where $H=F^TE$, $G = E^TE$, $F=B-P^{\left(\mathcal{S}\right)}B$, $E=A-P^{\left(\mathcal{S}\right)}A$ and $\mc{S}$ is the set of columns selected during the first $t-1$ iterations.
\end{problem}
For iteration $t$, define $\bs{\delta} = G_{:p}$, $\bs{\gamma} = H_{:p}$, $\bs{\omega} =
G_{:p}/\sqrt{G_{pp}} = \bs{\delta}/\sqrt{\bs{\delta}_{p}}$ and $\bs{\upsilon} =
H_{:p}/\sqrt{G_{pp}} = \bs{\gamma}/\sqrt{\bs{\delta}_{p}}$ \:. The vectors $\bs{\delta}^{(t)}$ and $\bs{\gamma}^{(t)}$ can be calculated in terms of $A$, $B$ and previous $\bs{\omega}$'s and $\bs{\upsilon}$'s
as
\begin{equation} \label{eq:delta_omega}
   \bs{\delta}^{(t)}=A^{T}A_{:p}-\sum_{r=1}^{t-1}\bs{\omega}_{p}^{(r)}\bs{\omega}^{(r)}, \:\:\:\:\:\: \bs{\gamma}^{(t)}=B^{T}A_{:p}-\sum_{r=1}^{t-1}\bs{\omega}_{p}^{(r)}\bs{\upsilon}^{(r)}\:.
\end{equation}
The numerator and denominator of the selection criterion at each iteration can be calculated in an efficient manner without explicitly calculating $H$ or $G$ using the following theorem. 
\begin{theorem} \label{Th:Rec_fg2}
Let $\bs{f}_{i}=\left\Vert H_{:i}\right\Vert ^{2}$ and
$\bs{g}_{i}=G_{ii}$ be the numerator and denominator of the
greedy criterion function for column $i$ respectively,
$\bs{f}=\left[\bs{f}_{i}\right]_{i=1..n}$, and
$\bs{g}=\left[\bs{g}_{i}\right]_{i=1..n}$. Then,
\begin{displaymath}
\begin{split}\bs{f}^{(t)}&=\Big(\bs{f}-2\left(\bs{\omega}\circ\left(A^{T}B\bs{\upsilon}-\Sigma_{r=1}^{t-2}\left(\bs{\upsilon}^{\left(r\right)T}\bs{\upsilon}\right)\bs{\omega}^{^{\left(r\right)}}\right)\right)
+\|\bs{\upsilon}\|^{2}\left(\bs{\omega}\circ\bs{\omega}\right)\Big)^{(t-1)},\\
\bs{g}^{(t)}
&=\Big(\bs{g}-\left(\bs{\omega}\circ\bs{\omega}\right)\Big)^{(t-1)}\:,\end{split}
\end{displaymath} where $\circ$ represents the Hadamard product operator.
\end{theorem}

In the update formulas of Theorem \ref{Th:Rec_fg2}, $A^TB$ can be calculated once and then used in different iterations. This makes the computational complexity of these formulas $\bigOO(nr)$ per
iteration. The computational
complexity of the algorithm is dominated by that of calculating
$A^TA_{:p}$ in (\ref{eq:delta_omega}) which is of $\bigOO(mn)$ per iteration. The other complex step is that of calculating the initial $\bs{f}$, which is $\bigOO(mnr)$. However, these steps can be implemented in an efficient way if the data matrix is sparse. The total computational complexity of the algorithm is $\bigOO(\max(mnl, mnr))$, where $l$ is the number of selected columns. Algorithm \ref{Alg:GenGCSS} in Appendix A shows the complete greedy algorithm.

\section{Generalized CSS Problems}
We describe a variety of problems that can be formulated as a generalized column subset selection (see Table \ref{tab:GCSS}). It should be noted that for some of these problems, the use of greedy algorithms has been explored in the literature. However, identifying the connection between these problems and the problem presented in this paper gives more insight about these problems, and allows the efficient greedy algorithm presented in this paper to be explored in other interesting domains.
\begin{table*}[t]
\caption{\label{tab:GCSS}Different problems as instances of the generalized column subset selection problem.}

\begin{center}

{\scriptsize }%
\begin{tabular}{ccc}
\hline
\textbf{\small Method} & \textbf{\small Source}{\small{} } & \textbf{\small Target }\tabularnewline
\hline
{\small Generalized CSS} & {\small $A$} & {\small $B$}\tabularnewline
\hline
{\small Column Subset Selection} & {\small Data matrix $A$} & {\small Data matrix $A$}\tabularnewline
{\small Distributed CSS} & {\small Data matrix $A$} & {\small Random subspace $A\Omega$ }\tabularnewline
{\small SVD-based CSS} & {\small Data matrix $A$} & {\small SVD-based subspace $U_{k}\Sigma_{k}$}\tabularnewline
{\small Sparse Approximation} & {\small Atoms $D$} & {\small Target vector $\bf{y}$}\tabularnewline
{\small Simultaneous Sparse Approximation} & {\small Atoms $D$} & {\small Target vectors $\left[\bf{y}_{\pth{1}}, \bf{y}_{\pth{2}}, ...  \bf{y}_{\pth{r}}\right]$}\tabularnewline
\hline
\end{tabular}{\scriptsize \par}

\end{center}
\end{table*}

\textbf{Column Subset Selection.} The basic column subset selection \cite{Frieze98-Rnd, Drineas06-Cols, Boutsidis08-Clust, Boutsidis09a-CSS, Boutsidis11a-NearOpt} is clearly an instance of the generalized CSS problem. In this instance, the target matrix is the same as the source matrix $B=A$ and the goal is to select a subset of columns from a data matrix that best represent other columns. The greedy algorithm presented in this paper can be directly used for solving the basic CSS problem. A detailed comparison of the greedy CSS algorithm and the state-of-the-art CSS methods can be found at \cite{Farahat12tt}. In our previous work \cite{farahat11-icdm, farahat12}, we successfully used the proposed greedy algorithm for unsupervised feature selection which is an instance of the CSS problem. We used the greedy algorithm to solve two instances of the generalized CSS problem: one is based on selecting features that approximate the original matrix $B=A$ and the other is based on selecting features that approximate a random partitioning of the features $B_{:c}=\sum_{j\in\mc{P}_{c}}A_{:j}$. The proposed greedy algorithms achieved superior clustering performance in comparison to state-of-the-art methods for unsupervised feature selection.


\textbf{Distributed Column Subset Selection.} The generalized CSS problem can be used to define distributed variants of the basic column subset selection problem. In this case, the matrix $B$ is defined to encode a concise representation of the span of the original matrix $A$. This concise representation can be obtained using an efficient method like random projection. In our recent work \cite{Farahat13css}, we defined a distributed CSS based on this idea and used the proposed greedy algorithm to select columns from big data matrices that are massively distributed across different machines.

\textbf{SVD-based Column Subset Selection.}
{\c{C}}ivril and Magdon-Ismail \cite{Civril12-CSS-Sparse} proposed a CSS method which first calculates the Singular Value Decomposition (SVD) of the data matrix, and then selects the subset of columns which best approximates the leading singular values of the data matrix. The formulation of this CSS method is an instance of the generalized CSS problem, in which the target matrix is calculated from the leading singular vectors of the data matrix. The greedy algorithm presented in \cite{Civril12-CSS-Sparse} can be implemented using Algorithm \ref{Alg:GenGCSS} by setting $B=U_{k}\Sigma_{k}$ where $U_{k}$ is a matrix whose columns represent the leading left singular vectors of the data matrix, and $\Sigma_{k}$ is a matrix whose diagonal elements represent the corresponding singular values. Our greedy algorithm is however more efficient than the greedy algorithm of \cite{Civril12-CSS-Sparse}.

\textbf{Sparse Approximation.} Given a target vector and a set of basis vectors, also called atoms, the goal of sparse approximation is to represent the target vector as a linear combination of a few atoms \cite{tropp2004greed}. Different instances of this problem have been studied in the literature under different names, such as variable selection for linear regression \cite{Das2008}, sparse coding \cite{olshausen1997sparse, lee2007efficient}, and dictionary selection \cite{CevherK11, DasK11}. If the goal is to minimize the discrepancy between the target vector and its projection onto the subspace of selected atoms, the sparse approximation can be considered an instance of the generalized CSS problem in which the target matrix is a vector and the columns of the source matrix are the atoms.
Several greedy algorithms have been proposed for sparse approximation, such as basic matching pursuit \cite{mallat1993matching}, orthogonal matching pursuit \cite{tropp2007signal}, the orthogonal least squares \cite{chen1989orthogonal}. 
The greedy algorithm for generalized CSS is equivalent to the orthogonal least squares algorithm (as defined in \cite{blumensath2007difference}) because at each iteration it selects a new column such that the reconstruction error after adding this column is minimum.  Algorithm \ref{Alg:GenGCSS} can be used to efficiently implement the orthogonal least squares algorithm by setting $B=\bf{y}$, where $\bf{y}$ is the target vector. However, an additional step will be needed to calculate the weights of the selected atoms as $\inv{A_{:\mc{S}}^{T}A_{:\mc{S}}} A_{:\mc{S}}^{T}\bf{y}$.

\textbf{Simultaneous Sparse Approximation.} A more general sparse approximation problem is the selection of atoms which represent a group of target vectors. This problem is referred to as simultaneous sparse approximation \cite{tropp2006algorithms}. Different greedy algorithms have been proposed for simultaneous sparse approximation with different constraints \cite{tropp2006algorithms, CevherK11}. If the goal is to select a subset of atoms to represent different target vectors without imposing sparsity constraints on each representation, simultaneous sparse approximation will be an instance of the greedy CSS problem, where the source columns are the atoms and the target columns are the input signals.

\section{Conclusions}
We define a generalized variant of the column subset selection problem and present a fast greedy algorithm for solving it. The proposed greedy algorithm can be effectively used to solve a variety of problems that are instances of the generalized column subset selection problem.

\newpage
\bibliographystyle{abbrv}

\bibliography{References}

\begin{thebibliography}{10}

\bibitem{blumensath2007difference}
T.~Blumensath and M.~E. Davies.
\newblock On the difference between orthogonal matching pursuit and orthogonal
  least squares.
\newblock 2007.
\newblock Unpublished Manuscript.

\bibitem{Boutsidis11a-NearOpt}
C.~Boutsidis, P.~Drineas, and M.~Magdon-Ismail.
\newblock Near optimal column-based matrix reconstruction.
\newblock In {\em Proceedings of the 52nd Annual IEEE Symposium on Foundations
  of Computer Science (FOCS'11)}, pages 305 --314, 2011.

\bibitem{Boutsidis09a-CSS}
C.~Boutsidis, M.~W. Mahoney, and P.~Drineas.
\newblock An improved approximation algorithm for the column subset selection
  problem.
\newblock In {\em Proceedings of the Twentieth Annual {ACM-SIAM} Symposium on
  Discrete Algorithms ({SODA}'09)}, pages 968--977, 2009.

\bibitem{Boutsidis08-Clust}
C.~Boutsidis, J.~Sun, and N.~Anerousis.
\newblock Clustered subset selection and its applications on it service
  metrics.
\newblock In {\em Proceedings of the Seventeenth {ACM} Conference on
  Information and Knowledge Management ({CIKM}'08)}, pages 599--608, 2008.

\bibitem{Civril12-CSS-Sparse}
A.~\c{C}ivril and M.~Magdon-Ismail.
\newblock {Column subset selection via sparse approximation of SVD}.
\newblock {\em Theoretical Computer Science}, 421(0):1 -- 14, 2012.

\bibitem{CevherK11}
V.~Cevher and A.~Krause.
\newblock Greedy dictionary selection for sparse representation.
\newblock {\em Journal of Selected Topics in Signal Processing}, 5(5):979--988,
  2011.

\bibitem{chen1989orthogonal}
S.~Chen, S.~A. Billings, and W.~Luo.
\newblock Orthogonal least squares methods and their application to non-linear
  system identification.
\newblock {\em International Journal of control}, 50(5):1873--1896, 1989.

\bibitem{Das2008}
A.~Das and D.~Kempe.
\newblock Algorithms for subset selection in linear regression.
\newblock In {\em Proceedings of the 40th Annual ACM Symposium on Theory of
  Computing (STOC'08)}, pages 45--54, 2008.

\bibitem{DasK11}
A.~Das and D.~Kempe.
\newblock Submodular meets spectral: Greedy algorithms for subset selection,
  sparse approximation and dictionary selection.
\newblock In {\em Proceedings of the 28th International Conference on Machine
  Learning, (ICML'11)}, pages 1057--1064, 2011.

\bibitem{Drineas06-Cols}
P.~Drineas, M.~Mahoney, and S.~Muthukrishnan.
\newblock Subspace sampling and relative-error matrix approximation:
  Column-based methods.
\newblock In {\em Approximation, Randomization, and Combinatorial Optimization.
  Algorithms and Techniques}, pages 316--326. Springer, 2006.

\bibitem{Farahat12tt}
A.~K. Farahat.
\newblock {\em {Greedy Representative Selection for Unsupervised Data
  Analysis}}.
\newblock PhD thesis, University of Waterloo, 2012.

\bibitem{Farahat13css}
A.~K. Farahat, A.~Elgohary, A.~Ghodsi, and M.~S. Kamel.
\newblock {Distributed column subset selection on MapReduce}.
\newblock In {\em Proceedings of the Thirteenth IEEE International Conference
  on Data Mining (ICDM'13)}, 2013.
\newblock In Press.

\bibitem{farahat11-icdm}
A.~K. Farahat, A.~Ghodsi, and M.~S. Kamel.
\newblock An efficient greedy method for unsupervised feature selection.
\newblock In {\em Proceedings of the Eleventh IEEE International Conference on
  Data Mining (ICDM'11)}, pages 161 --170, 2011.

\bibitem{farahat12}
A.~K. Farahat, A.~Ghodsi, and M.~S. Kamel.
\newblock Efficient greedy feature selection for unsupervised learning.
\newblock {\em Knowledge and Information Systems}, 35(2):285--310, 2013.

\bibitem{Frieze98-Rnd}
A.~Frieze, R.~Kannan, and S.~Vempala.
\newblock {Fast Monte-Carlo algorithms for finding low-rank approximations}.
\newblock In {\em Proceedings of the 39th Annual IEEE Symposium on Foundations
  of Computer Science (FOCS'98)}, pages 370 --378, 1998.

\bibitem{lee2007efficient}
H.~Lee, A.~Battle, R.~Raina, and A.~Ng.
\newblock Efficient sparse coding algorithms.
\newblock In {\em Advances in Neural Information Processing Systems 19
  (NIPS'06)}, pages 801--808. MIT, 2006.

\bibitem{lutkepohl1996handbook}
H.~L\"utkepohl.
\newblock {\em {Handbook of Matrices}}.
\newblock John Wiley \& Sons Inc, 1996.

\bibitem{mallat1993matching}
S.~Mallat and Z.~Zhang.
\newblock Matching pursuits with time-frequency dictionaries.
\newblock {\em Signal Processing, IEEE Transactions on}, 41(12):3397--3415,
  1993.

\bibitem{olshausen1997sparse}
B.~Olshausen and D.~Field.
\newblock {Sparse coding with an overcomplete basis set: A strategy employed by
  VI?}
\newblock {\em Vision Research}, 37(23):3311--3326, 1997.

\bibitem{tropp2004greed}
J.~Tropp.
\newblock Greed is good: Algorithmic results for sparse approximation.
\newblock {\em Information Theory, IEEE Transactions on}, 50(10):2231--2242,
  2004.

\bibitem{tropp2007signal}
J.~Tropp and A.~Gilbert.
\newblock Signal recovery from random measurements via orthogonal matching
  pursuit.
\newblock {\em Information Theory, IEEE Transactions on}, 53(12):4655--4666,
  2007.

\bibitem{tropp2006algorithms}
J.~Tropp, A.~Gilbert, and M.~Strauss.
\newblock {Algorithms for simultaneous sparse approximation. Part I: Greedy
  pursuit}.
\newblock {\em Signal Processing}, 86(3):572--588, 2006.

\end{thebibliography}

\newpage
\section*{Appendix A}
\begin{algorithm}[h]
\caption{\label{Alg:GenGCSS}Greedy Generalized
Column Subset Selection}
\textbf{Input:} Source matrix $A$, Target matrix $B$, Number of columns $l$\\
\textbf{Output:} Selected subset of columns $\mc{S}$

\begin{algorithmic}[1]
\STATE Initialize $\bs{f}_i^{(0)}=\|B^TA_{:i}\|^{2}$,
$\bs{g}_i^{(0)}=A_{:i}^TA_{:i}$ for $i=1\:...\:n$
\STATE Repeat $t=1\rightarrow l$:
\STATE \hspace{0.25cm} $p={\arg\max}_i\ \bs{f}_{i}^{(t)}/\bs{g}_{i}^{(t)}$,~~~$\mc{S}=\mc{S}\cup  \{p\}$
\STATE \hspace{0.25cm}
$\bs{\delta}^{(t)}=A^TA_{:p}-\sum_{r=1}^{t-1}\bs{\omega}_{p}^{(r)}\bs{\omega}^{(r)}$
\STATE \hspace{0.25cm} $\bs{\gamma}^{(t)}=B^TA_{:p}-\sum_{r=1}^{t-1}\bs{\omega}_{p}^{(r)}\bs{\upsilon}^{(r)}$
\STATE \hspace{0.25cm} $\bs{\omega}^{(t)}=\bs{\delta}^{(t)}/\sqrt{\bs{\delta}^{(t)}_{p}}$, $\bs{\upsilon}^{(t)}=\bs{\gamma}^{(t)}/\sqrt{\bs{\delta}^{(t)}_{p}}$
\STATE \hspace{0.25cm} Update $\bs{f}_i$'s, $\bs{g}_i$'s (Theorem \ref{Th:Rec_fg2})
\end{algorithmic}
\end{algorithm}

\subsection*{Proof of Theorem \ref{Th:Rec_fg2}}
Let $\mc{S}$ denote the set of columns selected during the
first $t-1$ iterations, $F^{(t-1)}$ denote the residual matrix of $B$ at the start of the $t$-th iteration (i.e., $F^{\left(t-1\right)}=B-P^{\left(\mathcal{S}\right)}B$), and $p$ be the column selected at iteration $t$. From Lemma \ref{Lm:Proj}, $P^{\left(\mathcal{S}\cup\left\{ p\right\} \right)}=P^{\left(\mathcal{S}\right)}+R^{\left(\left\{ p\right\} \right)}$. Multiplying both sides with $B$ gives $P^{\left(\mathcal{S}\cup\left\{ p\right\} \right)}B=P^{\left(\mathcal{S}\right)}B+R^{\left(\left\{ p\right\} \right)}B$. Subtracting both sides from $B$ and substituting $B-P^{\left(\mathcal{S}\right)}B$, and $B-P^{\left(\mathcal{S}\cup\left\{ p\right\} \right)}B$ with $F^{\left(t-1\right)}$ and $F^{\left(t\right)}$ respectively gives
$F^{\left(t\right)}=\left(F-R^{\left(\left\{ p\right\} \right)}B\right)^{\left(t-1\right)}.$

Since $R^{\left(\left\{ p\right\} \right)}B=R^{\left(\left\{ p\right\} \right)}F$ (see the proof of Theorem \ref{Th:RecF}), $F^{(t)}$ can be calculated recursively as
\begin{equation*} \label{Eq:ERec}
        F^{\left(t\right)}=\left(F-R^{\left(\left\{ p\right\} \right)}F\right)^{\left(t-1\right)}.
        \end{equation*}
Similarly, $E^{(t)}$ can be expressed as
        \begin{equation*}
       E^{\left(t\right)}=\left(E-R^{\left(\left\{ p\right\} \right)}E\right)^{\left(t-1\right)}.
    \end{equation*}
Substituting with $F$ and $E$ in $H=F^TE$ gives
\begin{equation*}
H^{\left(t\right)}=\left(\left(F-R^{\left(\left\{ p\right\} \right)}F\right)^{T}\left(E-R^{\left(\left\{ p\right\} \right)}E\right)\right)^{\left(t-1\right)}=\left(H-F^{T}R^{\left(\left\{ p\right\} \right)}E\right)^{\left(t-1\right)}.
\end{equation*}
Using $R^{\left(\left\{ p\right\} \right)}=E_{:p}\left(E_{:p}^{T}E_{:p}\right)^{-1}E_{:p}^{T}$, and given that $\boldsymbol{\omega}=G_{:p}=E^{T}E_{:p}/\sqrt{E_{:p}^{T}E_{:p}}$ and $\boldsymbol{\upsilon}=H_{:p}=F^{T}E_{:p}/\sqrt{E_{:p}^{T}E_{:p}}$, the matrix $H$ can be calculated recursively as
\begin{equation*}
  H^{\left(t\right)}=\left(H-\boldsymbol{\upsilon}\boldsymbol{\omega}^{T}\right)^{\left(t-1\right)}.
\end{equation*}
Similarly, $G$ can be expressed as
\begin{equation*}
  G^{\left(t\right)}=\left(G-\boldsymbol{\omega}\boldsymbol{\omega}^{T}\right)^{\left(t-1\right)}.
\end{equation*}
Using these recursive formulas, $\bs{f}_{i}^{(t)}$ can be
calculated as
\begin{equation*}
\begin{split}
\bs{f}_{i}^{\left(t\right)}&=\left(\left\Vert  H_{:i} \right\Vert ^{2}\right)^{(t)}=\left(\|H_{:i}-\bs{\omega}_i\bs{\upsilon}\|^{2}\right)^{\left(t-1\right)}\\
&=\left((H_{:i}-\bs{\omega}_i\bs{\upsilon})^T(H_{:i}-\bs{\omega}_i\bs{\upsilon})\right)^{\left(t-1\right)}\\
&=\left(H_{:i}^TH_{:i}-2\bs{\omega}_{i}H_{:i}^{T}\bs{\upsilon}+\bs{\omega}_{i}^{2}\|\bs{\upsilon}\|^{2}\right)^{\left(t-1\right)}\\
&=\left(\bs{f}_{i}-2\bs{\omega}_{i}H_{:i}^{T}\bs{\upsilon}+\bs{\omega}_{i}^{2}\|\bs{\upsilon}\|^{2}\right)^{\left(t-1\right)}.
\end{split}
\end{equation*}
Similarly, $\bs{g}_{i}^{(t)}$ can be calculated as
\begin{equation*}
\begin{split}
\bs{g}_{i}^{(t)}&=G_{ii}^{(t)}=\left(G_{ii}-\bs{\omega}_{i}^{2}\right)^{\left(t-1\right)}=\left(\bs{g}_{i}-\bs{\omega}_{i}^{2}\right)^{\left(t-1\right)}.
\end{split}
\end{equation*}
Let $\bs{f}=\left[\bs{f}_{i}\right]_{i=1..n}$and
$\bs{g}=\left[\bs{g}_{i}\right]_{i=1..n}$, $\bs{f}^{(t)}$ and
$\bs{g}^{(t)}$ can be expressed as
\begin{equation}\label{eq:f22}
\begin{split}
\bs{f}^{(t)}&=\left(\bs{f}-2\left(\bs{\omega}\circ
H^T\bs{\upsilon}\right)+\|\bs{\upsilon}\|^{2}\left(\bs{\omega}\circ\bs{\omega}\right)\right)^{(t-1)},
\\
\bs{g}^{(t)}&=\left(\bs{g}-\left(\bs{\omega}\circ\bs{\omega}\right)\right)^{(t-1)},
\end{split}
\end{equation}
where $\circ$ represents the Hadamard product operator.

Using the recursive formula of $H$, the
term $H^T\bs{\upsilon}$ at iteration $(t-1)$ can be expressed as
\begin{equation*}
\begin{split}
H^T\bs{\upsilon}&=\left(A^TB-\Sigma_{r=1}^{t-2}\left(\bs{\omega}\bs{\upsilon}^{T}\right)^{\left(r\right)}\right)\bs{\upsilon}
=A^{T}B\bs{\upsilon}-\Sigma_{r=1}^{t-2}\left(\bs{\upsilon}^{\left(r\right)T}\bs{\upsilon}\right)\bs{\omega}^{^{\left(r\right)}}
\end{split}
\end{equation*}
Substituting with $H^T\bs{\upsilon}$ in (\ref{eq:f22}) gives the update formulas for $\bs{f}$ and $\bs{g}$. \hfill\BlackBox

\end{document}